\documentclass[
reprint,           % journal's actual layout
superscriptaddress,% authors with affiliations via superscripts
amsmath,           % add AMS-Latex features
amssymb,           % add extra AMS symbols, including amsfonts
aps,               % aps or aip
prl,               % prl, pra, prb, prc, prd, pre, prstab
notitlepage,       % control appearance of title page
longbibliography,  % show  article titles in the bibliography
floatfix,          % process floats as early as possible
nofootinbib,
]{revtex4-1}

\usepackage{tensor}     % manipulate tensors
\usepackage{graphicx}   % include figures
\usepackage[
colorlinks=true,        % color link
citecolor=blue,         % cite color
linkcolor=blue,         % link color
urlcolor=blue           % url color
]{hyperref}             % create hyperlinks
\usepackage{bm}         % \bm{<text>} Bold math symbols
\usepackage{xcolor}     % textcolor
\usepackage{array}      % dcolumn depends on array
\usepackage{lipsum}
\usepackage{color}      % \textcolor{declared-color}{text}
\usepackage[utf8]{inputenc} % accented characters for .bib file using XeLaTex
\usepackage[section]{placeins} % figures processing
\usepackage{multirow}

\newcommand{\nc}{\newcommand*} 

\nc{\al}{\alpha}
\nc{\s}{\sigma}
\nc{\kp}{\kappa}
\nc{\dt}{\delta}
\nc{\Dt}{\Delta}
\nc{\Ld}{\Lambda}
\nc{\p}{\partial}
\nc{\Gm}{\Gamma}
\nc{\om}{\omega}
\nc{\Om}{\Omega}
\nc{\rd}{\mathrm{d}}
\nc{\Od}{\mathcal{O}} % order operator
\def\({\left(}
\def\){\right)}
\def\[{\left[}
\def\]{\right]}
\def\e{\begin{equation}}
\def\q{\end{equation}}
\def\m{\begin{eqnarray}}
\def\n{\end{eqnarray}}
\nc{\Eq}[1]{Eq.~\eqref{#1}}     % equation
\nc{\Fig}[1]{Fig.~\ref{#1}}     % figure
\nc{\Table}[1]{Table~\ref{#1}}  % table
\nc{\Sec}[1]{Sec.~\ref{#1}}     % section
\nc{\Msun}{M_\odot}             % solar mass
\nc{\fpbh}{f_{\mathrm{PBH}}}    % f_pbh
\nc{\fpbhn}{f_{\mathrm{pbh0}}}    % f_pbh
\nc{\mR}{\mathcal{R}} % merger rate density
\nc{\seq}{\sigma_{\mathrm{eq}}}
\nc{\ogw}{\Omega_{\mathrm{GW}}}
\nc{\gpcyr}{\mathrm{Gpc}^{-3}\,\mathrm{yr}^{-1}}
\nc{\lvc}{LIGO-Virgo} % LIGO-VIRGO collaboration
\nc{\SNR}{\mathrm{SNR}} % signal to noise ratio
\nc{\mmin}{{m_{\mathrm{min}}}}
\nc{\mmax}{{m_{\mathrm{max}}}}
\nc{\Mmin}{{M_{\mathrm{min}}}}
\nc{\fmin}{{f_{\mathrm{min}}}}
\nc{\VT}{\mathrm{VT}}
\nc{\rhoGW}{\rho_{\mathrm{GW}}}
\nc{\vth}{\vec{\theta}}
\nc{\vd}{\vec{d}}
\nc{\vla}{\vec{\lambda}}
\nc{\Nobs}{N_{\mathrm{obs}}}
\nc{\av}[1]{\langle #1 \rangle} % average bracket
\nc{\km}{\mathrm{km}}
\nc{\Mpc}{\mathrm{Mpc}}
\nc{\Tobs}{T_{\mathrm{obs}}}
\nc{\Ntemp}{N_{\mathrm{temp}}}
\nc{\fyr}{f_{\mathrm{yr}}}
\nc{\addref}{[\textcolor{red}{add ref}] } % placeholder of references
\nc{\eg}{\textit{e.g.~}}
\nc{\app}{\approx}
\nc{\hf}{\frac{1}{2}}
\nc{\discuss}{\textcolor{red}{Add discussion here!}}
\nc{\red}[1]{\textcolor{red}{#1}}
\newcommand{\dif}{\mathrm{d}}

\begin{document}
	
%%%%%%%%%%%%%%%%%%%%%%%%%%%%%%%%%%%% title %%%%%%%%%%%%%%%%%%%%%%%%%%%%%%%%%%%%%
\title{Constraining the Gravitational-Wave Spectrum from Cosmological First-Order Phase Transitions Using Data from LIGO-Virgo First Three Observing Runs}
	
%%%%%%%%%%%%%%%%%%%%%%%%%%%%%%%%%%%% author %%%%%%%%%%%%%%%%%%%%%%%%%%%%%%%%%%%%
\author{Yang Jiang}
\email{jiangyang@itp.ac.cn} 
\affiliation{CAS Key Laboratory of Theoretical Physics, 
		Institute of Theoretical Physics, Chinese Academy of Sciences,
		Beijing 100190, China}
\affiliation{School of Physical Sciences, 
		University of Chinese Academy of Sciences, 
		No. 19A Yuquan Road, Beijing 100049, China}
	
	%%%%%%%%%%%%%%%%%%%%%%%%%%%%%%%%%%%% author %%%%%%%%%%%%%%%%%%%%%%%%%%%%%%%%%%%%
\author{Qing-Guo Huang}
\email{Corresponding author: huangqg@itp.ac.cn}
\affiliation{CAS Key Laboratory of Theoretical Physics, 
		Institute of Theoretical Physics, Chinese Academy of Sciences,
		Beijing 100190, China}
\affiliation{School of Physical Sciences, 
		University of Chinese Academy of Sciences, 
		No. 19A Yuquan Road, Beijing 100049, China}
\affiliation{School of Fundamental Physics and Mathematical Sciences
		Hangzhou Institute for Advanced Study, UCAS, Hangzhou 310024, China}
%\affiliation{Center for Gravitation and Cosmology,  College of Physical Science and Technology, Yangzhou University, Yangzhou 225009, China}
%\affiliation{Shanghai Frontier Science Research Center for Gravitational Wave Detection, School of Aeronautics and Astronautics, Shanghai Jiao Tong University, Shanghai 200240, China}

	%%%%%%%%%%%%%%%%%%%%%%%%%%%%%%%%%%%%% date %%%%%%%%%%%%%%%%%%%%%%%%%%%%%%%%%%%%%
\date{\today}
	
	%%%%%%%%%%%%%%%%%%%%%%%%%%%%%%%%% abstract %%%%%%%%%%%%%%%%%%%%%%%%%%%%%%%%%%%%%
\begin{abstract}
We search for a first-order phase transition (PT) gravitational wave (GW) signal from Advanced LIGO and Advanced Virgo's first three observing runs. Due to the large theoretical uncertainties, four GW energy spectral shapes from bubble and sound wave collisions widely adopted in literature are investigated, separately. Our results indicate that there is no evidence for the existence of such GW signals, and therefore we give the upper limits on the amplitude of GW energy spectrum $\Omega_\text{pt}(f_*)$ in the peak frequency range of $f_*\in [5,500]$ Hz for these four theoretical models, separately. We find that $\Omega_\text{pt}(f_*\simeq 40\ \text{Hz})<1.3\times10^{-8}$ at $95\%$ credible level, and roughly  $H_*/\beta\lesssim 0.1$ and $\alpha\lesssim 1$ at $68\%$ credible level in the peak frequency range of $20\lesssim f_*\lesssim 100$ Hz corresponding to the most
sensitive frequency band of Advanced LIGO and Advanced Virgo's first three observing runs, where $H_*$ is the Hubble parameter when PT happens, $\beta$ is the bubble nucleation rate and $\alpha$ is the ratio of vacuum and relativistic energy density. 

\end{abstract}
	
\maketitle
%%%%%%%%%%%%%%%%%%%%%%%%%%%%%%%%%%%%%%%%%%%%%%%%%%%%%%%%%%%%%%%%%%%%%%%%%%%%%%%%
{\it Introduction.} 
The evolution of the Universe can be carried out in a smooth or abrupt way, and these sudden changes are what we called phase transitions (PTs) \cite{Linde_1979,KIBBLE1980183,Mazumdar:2018dfl}.
 The early evolution of the Universe might contain a variety of PTs since many extensions of Standard Model, such as grand unification model \cite{Croon:2019kpe}, suppersymmetric model \cite{Huber:2015znp} and so on, predict the occurrence of PTs. See some other relevant models in  \cite{Megias:2018sxv,Fromme:2006cm,Huang:2016cjm,Hebecker:2016vbl,Jinno:2016knw,Brdar:2019fur,Cai:2022bcf,Hall:2019rld,Hall:2019ank}. At QCD scale, PT could happen due to the confinement between quarks and gluons \cite{Schwarz:2009ii,PhysRevLett.105.041301}. At a higher energy scale, the destruction of electroweak gauge symmetry may lead to electroweak PT \cite{Kajantie:1995kf}. Generally, each breaking of new symmetry introduced by theoretical models might lead to the occurrence of PT. Therefore, the observation of PT becomes a way to explore new physics.

Among all kinds of PTs, the first-order PT is of great significance for gravitational wave (GW) astronomy, because they are usually strong enough to be interesting sources of GW radiation \cite{PhysRevD.30.272,10.1093/mnras/218.4.629} in the Universe. When the temperature of the Universe drops to a certain value, bubbles containing true vacuum will nucleate in the meta-stable vacuum. Subsequently, these bubbles expand, merge, fill the Universe in the end and GWs are produced during this process \cite{PhysRevLett.69.2026,PhysRevD.45.4514,Kamionkowski:1993fg}. Since the process of nucleation and collision occurs in a random manner, the radiated GWs form a stochastic gravitational wave background (SGWB). Because the gravitational interaction is very weak, SGWB decouples from the primordial plasma rapidly. Detecting these GW signals has become an important way to explore the early Universe which is difficult to be observed by any other methods due to the longer decoupling time \cite{Maggiore:1999vm}. 

At the time when GWs are produced, its frequency depends on the duration of PT, and then GWs would be redshifted by the expansion of the Universe. Therefore, the frequency band of SGWB today is also related to the temperature $T_*$ at which PT occurs. Pulsar Timing Arrays (PTAs) \cite{Brazier:2019mmu,Desvignes:2016yex,Kerr:2020qdo,Perera:2019sca} are used to search for the SGWB at  frequencies of several nHz, corresponding $T_*$ at the order of MeV. Even though a stochastic process has been detected by PTA data sets \cite{NANOGrav:2020bcs,Goncharov:2021oub,Chen:2021ncc,Antoniadis:2022pcn}, a SGWB detection consistent with general relativity cannot be claimed because there is no statistically significant evidence of quadrupolar spatial correlations \cite{Chen:2021wdo,Wu:2021kmd,NANOGrav:2021ini}, and the constraints on the first-order PT parameters from PTA data sets are presented in  \cite{NANOGrav:2021flc,Xue:2021gyq}.
%The observation result of NANOGrav program has shown signs of stochastic process \cite{NANOGrav:2020bcs} and it might be expressed as a signal from PT \cite{Li:2021qer,Ratzinger:2020koh,NANOGrav:2021flc}. 
On the other hand, the terrestrial GW observatories like Advanced LIGO \cite{LIGOScientific:2014pky} and Advanced Virgo \cite{VIRGO:2014yos} are designed to be sensitive to the frequency of $10\sim10^3$ Hz roughly corresponding to the PT temperature in the range of $10^5\sim10^{10}$ GeV. So far, LIGO/Virgo/KAGRA Collaboration has accumulated three generations (O1-O3) of GWs observing data and there is no evidence for the SGWB signal \cite{KAGRA:2021kbb,KAGRA:2021mth}. The related constraints on the GW energy spectrum generated by the first order PT have been investigated in \cite{Romero:2021kby} based on the broken power law and two phenomenological PT models.

In this letter, we present a comprehensive analysis of the first-order PT models utilizing data from Advanced LIGO and Advanced Virgo's first three observing runs by extending and improving the analysis in \cite{Romero:2021kby} in several aspects. First of all, we take into account four different GW energy spectral shapes from bubble and sound wave collisions widely adopted in literature due to the large theoretical uncertainties, and provide the constraints on the amplitude of the GW energy spectrum in the peak frequency range of $f_*\in [5, 500]$ Hz for these four theoretical models, separately. Secondly, although compact binary coalescences (CBCs) background is supposed to contribute SGWB as well, it is not strong enough to produce a detectable correlation according to the current sensitivities of both Advanced LIGO and Advanced Virgo \cite{KAGRA:2021kbb}. Therefore, different from \cite{Romero:2021kby}, we do not include the contribution of CBCs in our analysis. Thirdly, the parameters related to the amplitude of the GW spectrum are chosen to be uniform distributions in our analysis for resulting in more conservative upper limits than the log-uniform priors adopted in \cite{Romero:2021kby}. Fourthly, we do not keep the bubble nucleation rate $\beta$ and the PT temperature $T_*$ fixed for deriving the upper limits on the GW spectrum because they should be free from the viewpoint of data analysis.

%Finally, we scan the frequency domain and report the constraint capacity of aLIGO and aVirgo for different frequencies of PT backgrounds.
\smallskip

{\it SGWB from the first-order PT.}
Isotropic SGWB can be described by the fractional energy density spectrum in the frequency domain:
\begin{equation}
    \Omega(f)=\frac{1}{\rho_\text{c}}\frac{\dif\rho_\text{gw}}{\dif \ln f},
\end{equation}
where $\rho_\text{c}=3H_0^2c^2/(8\pi G)$ denotes the critical density of the current universe. It has been known that there are three main sources of GW produced by a first-order PT: bubble collisions, collisions of sound waves and magnetohydrodynamic turbulence \cite{Hindmarsh:2020hop,Weir:2017wfa}. In this letter, the contribution from turbulence is not considered due to the lack of understanding about its energy spectrum \cite{Caprini:2015zlo,RoperPol:2019wvy,PhysRevD.78.123006,PhysRevD.81.023004}. Besides, magnetohydrodynamic turbulence is always subdominant compared with sound waves. The amplitude of the GW energy spectrum from a first-order PT is characterized by the bubble nucleation rate $\beta$ and the ratio of vacuum and relativistic energy density $\alpha$. Quantitatively, the following energy density spectrum can be used to fit the SGWB from both bubble collisions and sound waves \cite{Caprini:2015zlo,Binetruy:2012ze,Jinno:2016vai,Hindmarsh:2017gnf,NANOGrav:2021flc}:
\begin{equation}
    h^2\Omega_\text{pt}(f)=\mathcal{R}g_*^{-\frac{1}3}\Delta(v_w)\left(\frac{\kappa\alpha}{1+\alpha}\right)^p\left(\frac{H_*}{\beta}\right)^q\mathcal{S}\left(\frac{f}{f_*}\right), \label{eq:template}
\end{equation}
where $h$ is the dimensionless Hubble constant, the factor $\mathcal{R}\simeq7.69\times10^{-5}$, $\Delta(v_w)$ is a function of bubble walls velocity $v_w$, $g_*$ is the number of relativistic degrees of freedom which is fixed to be $100$ in this letter, $H_*$ is the Hubble parameter when phase transition happens and $\kappa$ counts for the fraction of vacuum energy converted. $\mathcal{S}(x)$ represents the shape of the spectrum and it comes to its maximum value at $x=1$. Here, $f_*$ is the peak frequency of SGWB at present:
\begin{equation}
    f_*\simeq1.13\times10^{-7}\left(\frac{\tilde{f_*}}{\beta}\right)\left(\frac{\beta}{H_*}\right)\left(\frac{T_*}{\text{GeV}}\right)\left(\frac{g_*}{10}\right)^{1/6}\text{Hz}, \label{eq:f_star0}
\end{equation}
where $\tilde{f_*}$ is the peak frequency when PT happens. For sound waves, a suppression factor $\Upsilon$ should be multiplied in \Eq{eq:template} counting for the finite lifetime \cite{Ellis:2020awk,Guo:2020grp}:
\begin{equation}
    \Upsilon=1-(1+2\tau_\text{sw}H_*)^{-1/2}.
\end{equation}
The time scale $\tau_\text{sw}$ is usually taken to be the timescale for the onset of turbulence \cite{Weir:2017wfa}: $\tau_\text{sw}\simeq R_*/\bar{U}_f$, where $R_*=(8\pi)^{1/3}\beta^{-1}\mathrm{Max}(v_w,c_s)$ \cite{Hindmarsh:2019phv,Guo:2020grp} and $\bar{U}_f^2\simeq3\kappa_\text{sw}\alpha/[4(1+\alpha)]$ \cite{Weir:2017wfa}. More details about the  parameters in \Eq{eq:template} and \Eq{eq:f_star0} are listed in \Table{tab:template}. 

\begin{table}[ht]
    \centering
    \begin{tabular}{ccc}
        \hline\hline
         & Bubbles Collisions & Sound Waves \\
        \hline
        $\Delta(v_w)$ & $\frac{0.48v_w^3}{1+5.3v_w^2+5v_w^4}$ & $0.513v_w$ \\
        $\kappa$ & $1$ & $f(\alpha,v_w)$ \\
        $p$ & $2$ & $2$ \\
        $q$ & $2$ & $1$ \\
        $\mathcal{S}(x)$ & $\frac{(a+b)^c}{\left(bx^{-a/c}+ax^{b/c}\right)^c}$ & $x^3\left(\frac{7}{4+3x^2}\right)^{7/2}$ \\
        $\tilde{f_*}/\beta$ & $\frac{0.35}{1+0.07v_w+0.69v_w^4}$ & $\frac{0.536}{v_w}$\\
        \hline\hline
    \end{tabular}
    \caption{Parameters for the GW energy spectrum in \Eq{eq:template} and \Eq{eq:f_star0}}
    \label{tab:template}
\end{table}

\begin{figure}[ht]
    \centering
    \includegraphics[width=\columnwidth]{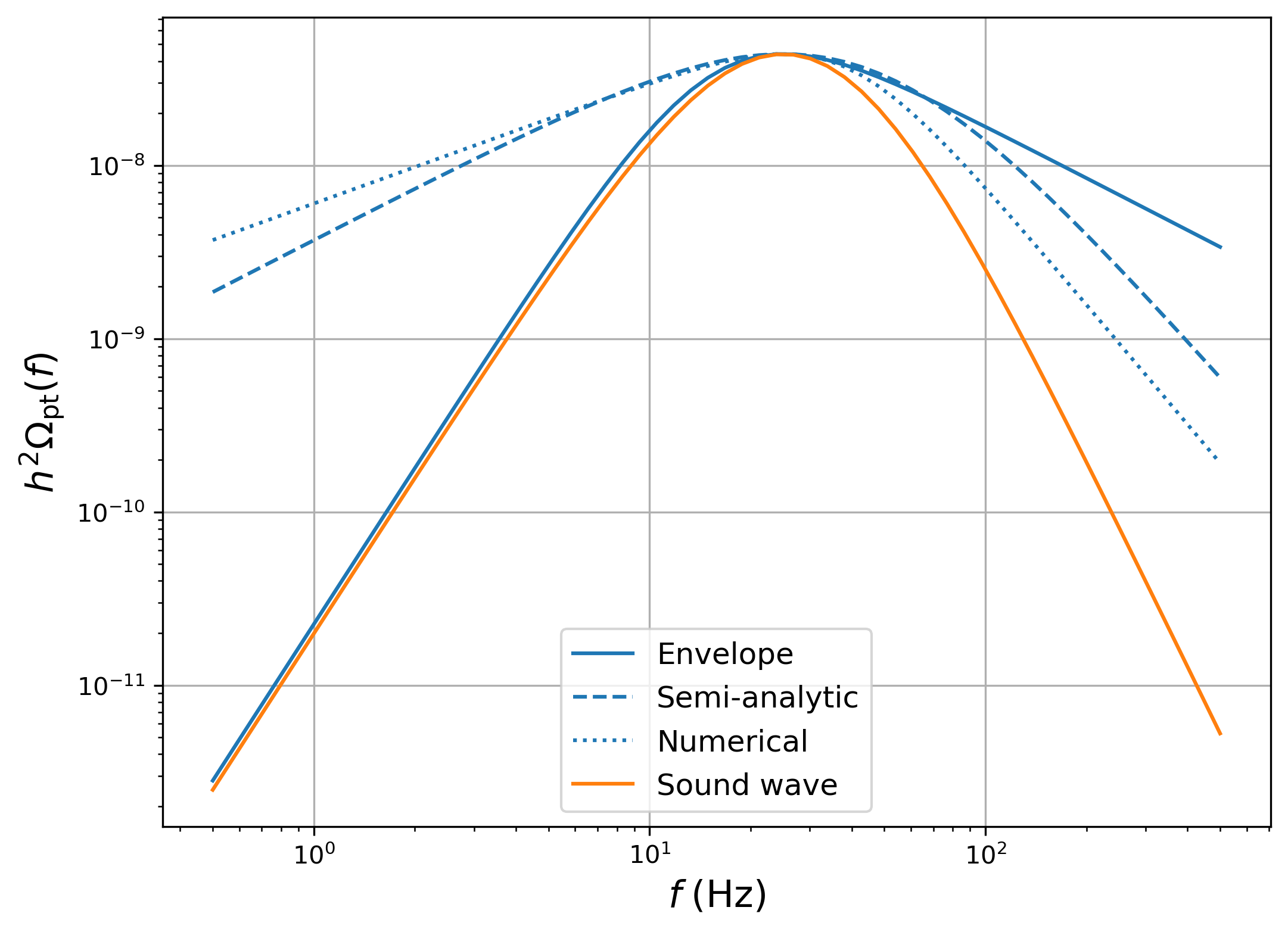}
    \caption{SGWB energy spectra for bubble collision and sound wave. The peak frequency $f_*$ is chosen to be $25$ Hz.}
    \label{fig:template}
\end{figure}

In fact, there are still large theoretical uncertainties for predicting the PT GW energy spectrum. Analytically, GW energy spectrum can be calculated assuming that energy is concentrated on the infinitesimal bubble wall and it vanishes once the bubbles collide with others. This method is called envelope approximation \cite{Jinno:2016vai,PhysRevD.45.4514}. Numerically, 3D lattice simulations \cite{Cutting:2020nla,Cutting:2018tjt} can be used to break through these assumptions. However, large solved volume to accomodate multiple bubbles and very dense lattices to fit thin bubble walls usually lead to substantial costs. Hence, semi-analytic \cite{Lewicki:2020azd} methods are served as alternative ways to do the calculation. \Table{tab:bubbletemplate} illustrates the details of different GW energy spectra and the typical shapes of these spectra are shown in \Fig{fig:template}. In the high frequency band, the numerical simulation results show faster attenuation than envelope approximation, while the opposite is true at low frequency band.
\begin{table}[ht]
    \centering
    \begin{tabular}{cccc}
        \hline\hline
         & Envelope & Semi-analytic & Numerical  \\
         \hline
         $a$ & $3$ & $1\sim2.3$ & $1.6\sim0.7$ \\
         $b$ & $1$ & $2.2\sim2.4$ & $1.4\sim2.3$ \\
         $c$ & $1.5$ & $2\sim4.2$ & $1$ \\
         $\tilde{f_*}/\beta$ & $\frac{0.35}{1+0.07v_w+0.69v_w^4}$ & $0.1$ & $0.2$ \\
         \hline\hline
    \end{tabular}
    \caption{Energy spectra of bubble collisions for envelope approximation, semi-analytic approach and lattice simulations. In this letter, we take $(a,b,c)=(1,2.2,2)$ and $(a,b,c)=(0.7,2.3,1)$ for semi-analytic and numerical methods respectively.}
    \label{tab:bubbletemplate}
\end{table}
\smallskip

{\it Method of data analysis.}
The measurement of SGWB depends on the correlations \cite{Romano:2016dpx,PhysRevD.59.102001} between multiple detectors and Bayesian approach is used to calculate the possibility of models \cite{PhysRevLett.109.171102}. Correlated magnetic noise budget shows that such signal is negligible \cite{LIGOScientific:2019vic,KAGRA:2021kbb} since the intensity of this correlation is much lower than the sensitivity of the current detectors. Therefore, a Gaussian distributed fluctuation is assumed and likelihood is given by 
\begin{equation}
     p(\hat{C}_{IJ}|\bm{\theta};\lambda)\propto \exp\left[-\frac12\sum_{IJ}\sum_{f}\frac{\left(\hat{C}_{IJ}(f)-\lambda\Omega(f;\bm{\theta})\right)^2}{\sigma^2_{IJ}(f)}\right].
\end{equation}
The energy spectrum is characterized by parameters $\bm{\theta}$ to be estimated. $\hat{C}_{IJ}$ denotes the spectrum of correlation and $\sigma_{IJ}$ is related to the noise intensity of detector pair $IJ$. Summing the index $IJ$ means multiplying the likelihoods from all detector pairs since the correlations between different baselines can be neglected \cite{PhysRevD.59.102001}. Here $\lambda$ denotes the calibration uncertainties \cite{Sun_2020} of the detector baselines and should be marginalized by the method given in \cite{Whelan_2014}.

The Bayesian analyses for the circumstances of bubble collision dominant and sound wave dominant cases will be done separately because it is ambiguous to determine which one plays a leading role. Python software {\it bilby} \cite{Ashton_2019} is used for parameter estimation and generate 2D posterior distribution corner-plots. For the bubble collision dominant case, all three models are considered, and $\kappa$ and $v_w $ are set to unity based on the assumption that the bubble wall interacts weakly with the plasma and runaway regime has been reached. For the sound wave dominant case, $\kappa$ is related to $\alpha$ and $v_w$ \cite{Espinosa:2010hh}. 
In our statistical analysis, $\bm{\theta}=(\alpha,\,H_*/\beta,\,f_*)$ are free parameters to be determined in the bubble collision dominant cases, and $\bm{\theta}=(\alpha,\,H_*/\beta,\,f_*,\,v_w)$ are free parameters in the sound wave dominant case. The parameters related to the amplitude of the GW energy spectrum are chosen to be uniform distributions in our analysis because it results in more conservative upper bounds than the log-uniform prior. Besides, we note that \Eq{eq:template} may not be applicable for $\alpha\gtrsim10$ and $H_*/\beta\gtrsim1$ \cite{Ellis:2019oqb,Ellis:2018mja}. Details about the priors are listed in \Table{tab:priors}.

\begin{table}[ht]
    \centering
    \begin{tabular}{cc}
        \hline\hline
         Parameter & Prior \\
         \hline
         $\alpha$ &  $\text{Uniform}(10^{-3},\,10)$ \\
         $H_*/\beta$ & $\text{Uniform}(10^{-3},\,1)$ \\
         $v_w$ & $\text{Uniform}(10^{-2},\,1)$ \\
         $f_*$ & $\text{LogUniform}(5,\,500)$ \\
         \hline\hline
    \end{tabular}
    \caption{Priors distributions of the parameters used for Bayesian analysis, and the distribution of $v_w$ is set to be $\delta(1)$ for the bubble collision dominant cases.}
    \label{tab:priors}
\end{table}
\smallskip

{\it Results.}
\begin{figure*}[ht]
    \centering
    \includegraphics[width=\columnwidth]{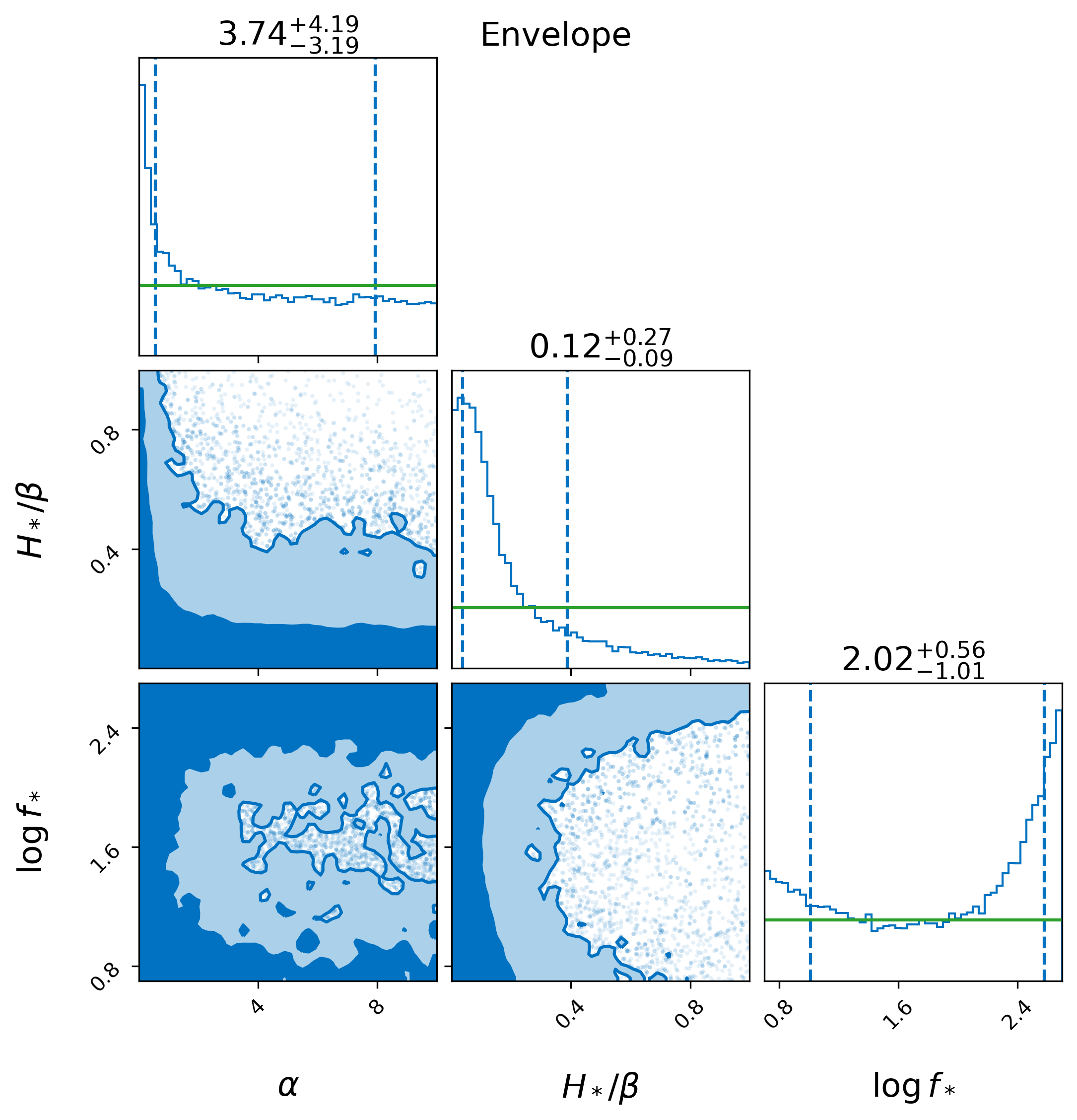}
    \includegraphics[width=\columnwidth]{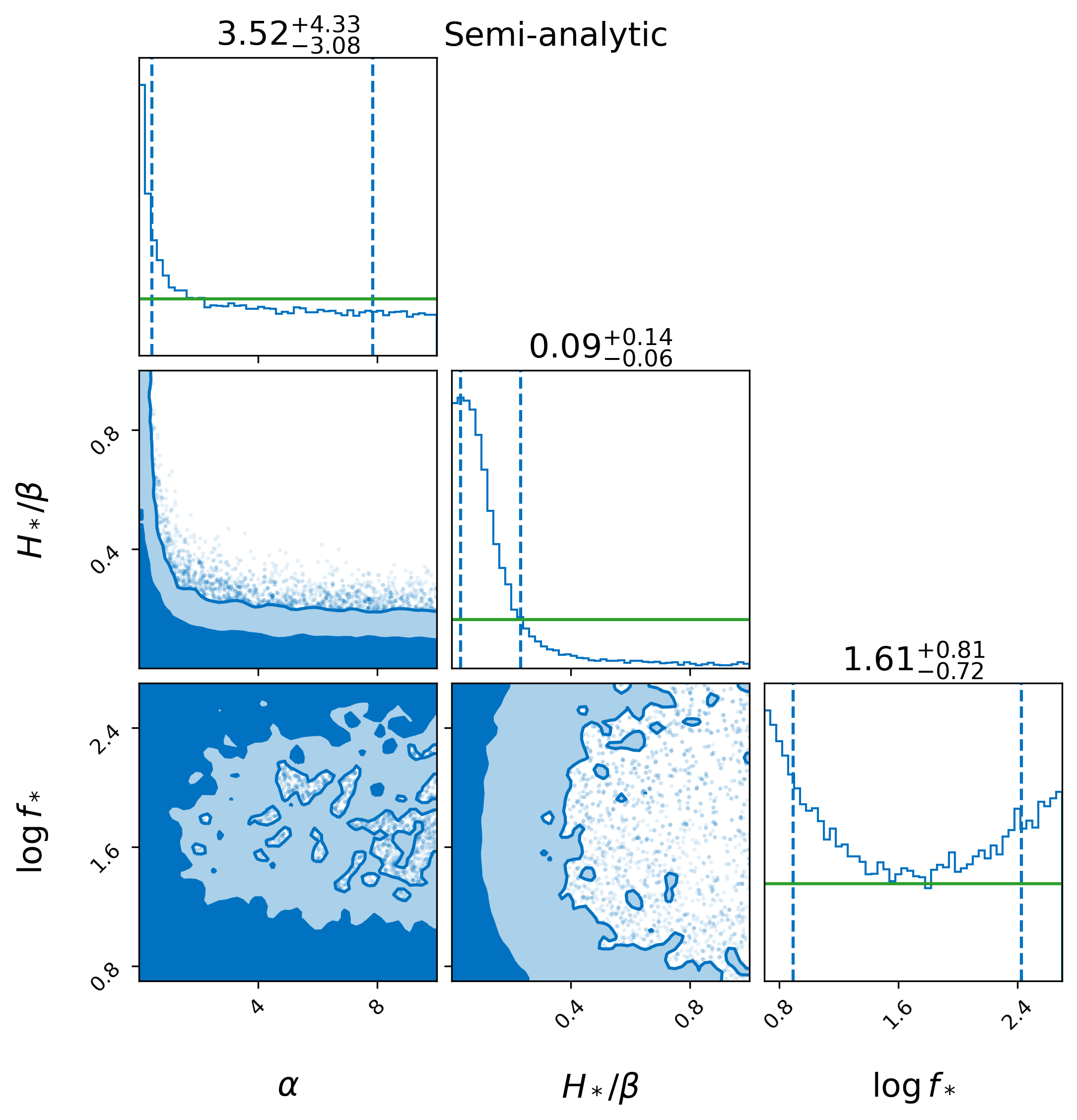} \\
    \includegraphics[width=\columnwidth]{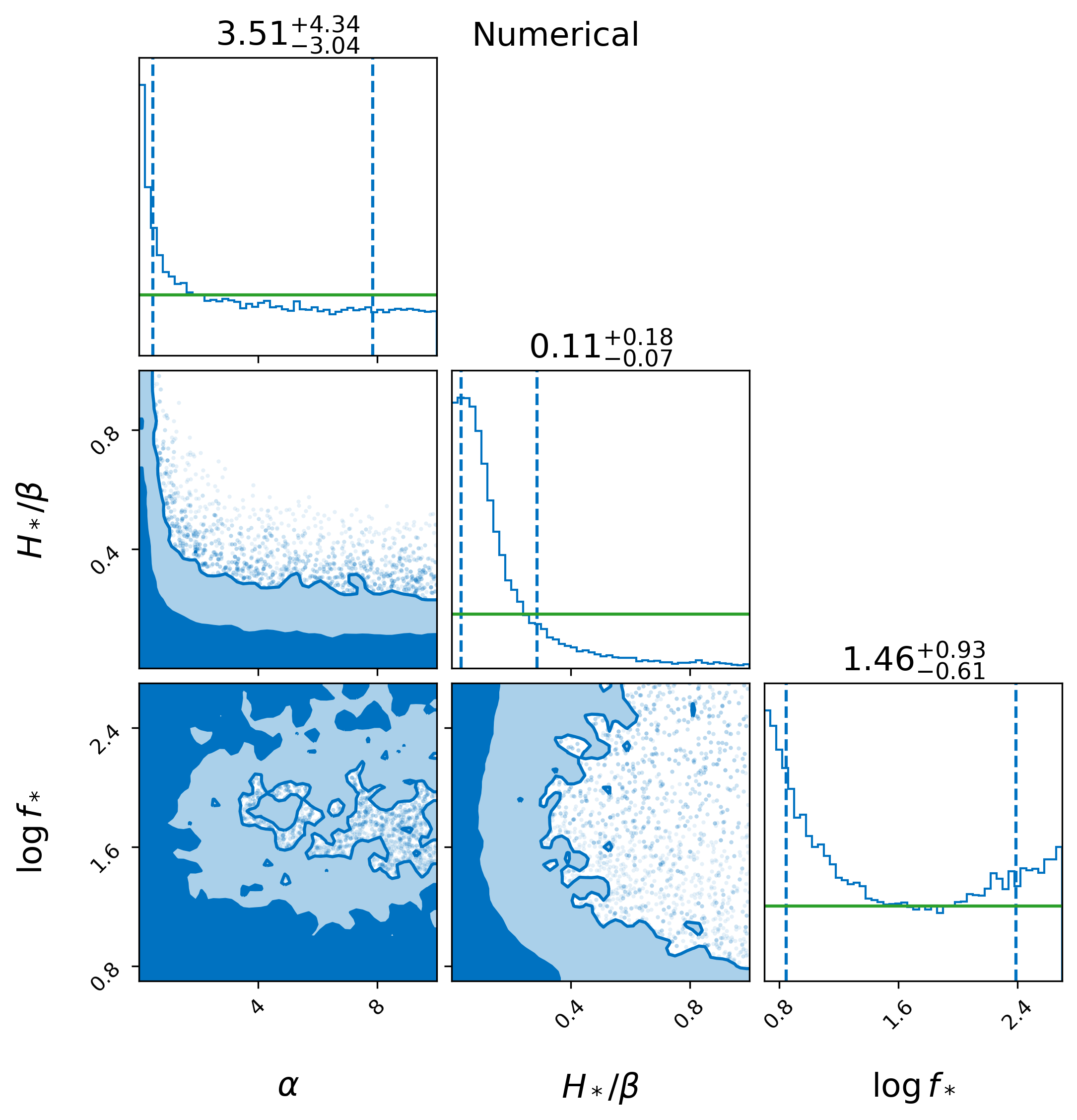}
    \includegraphics[width=\columnwidth]{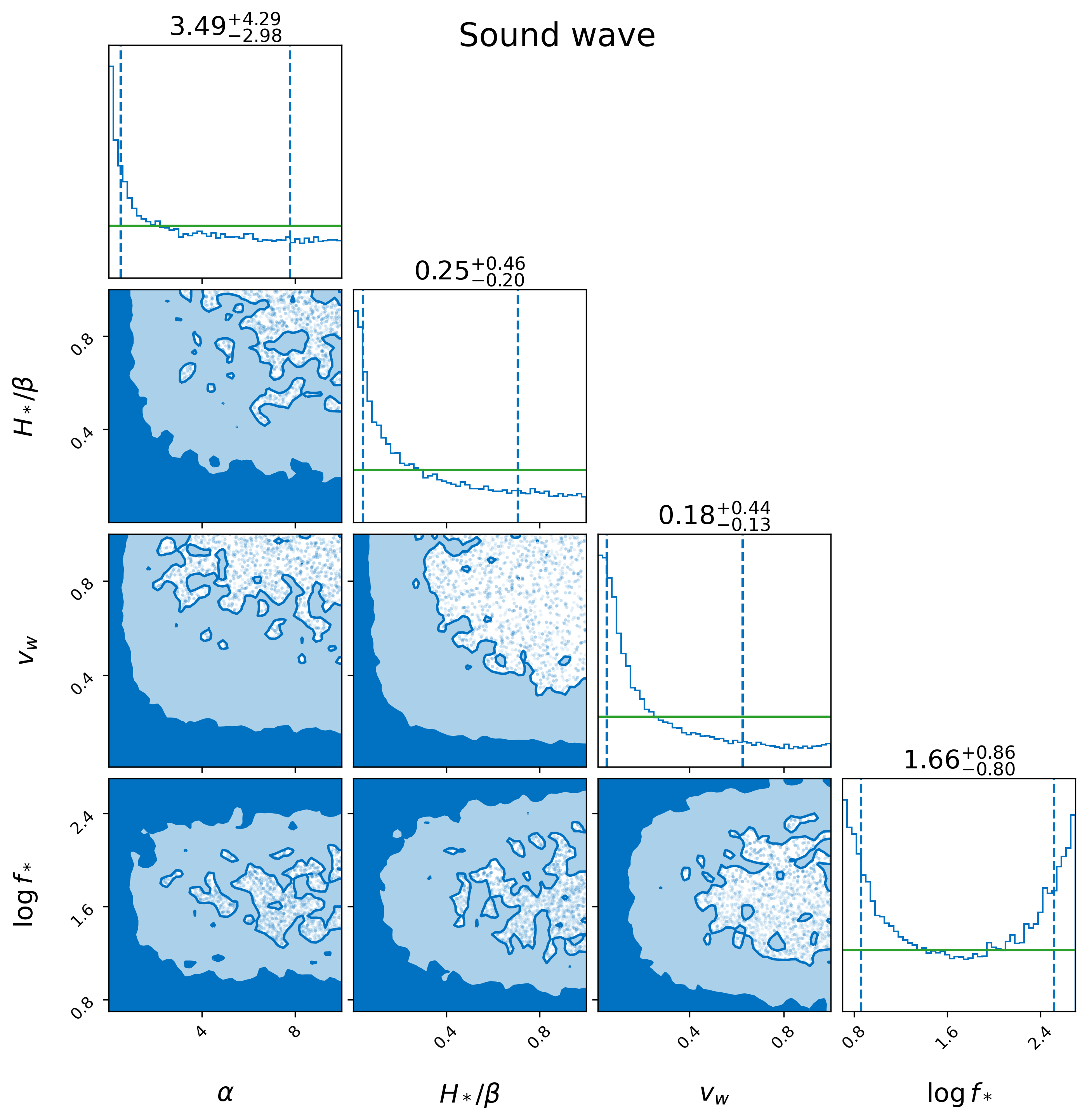}
    \caption{Posterior distributions of parameters for bubble collision and sound wave dominant cases. Here $68\%$ and $95\%$ exclusion contours are shown. Horizontal solid green lines denote the priors used in analysis.}
    \label{fig:cornerplot}
\end{figure*}
The posterior distributions of parameters are illustrated in \Fig{fig:cornerplot} for all four theoretical models. Roughly speaking, we find $H_*/\beta\gtrsim0.1$ and $\alpha\gtrsim 1$ are excluded at $68\%$ confidence level in the peak frequency range of $f_*\in [20, 100]$ Hz for all four theoretical models considered in this letter. This frequency range rougly corresponds to the most sensitive frequency band of Advanced LIGO and Advanced Virgo since $99\%$ of the sensitivity measuring the SGWB comes from the frequency band of $20\sim76.6\,\text{Hz}$ during O3 observation run \cite{KAGRA:2021kbb}. The Bayes factors between SGWB caused by PT and pure noise $\log\mathcal{B}_\text{niose}^\text{PT}$ are $-0.64$, $-0.74$, $-0.70$ and $-0.63$ for the bubble collision dominant cases fitted by envelope approximation, semi-analytic, numerical methods, and the sound wave dominant case, respectively. This result indicates that there is no evidence to claim such SGWB signals in the data. 
%Therefore, we provide the upper limits of the energy spectrum within the posterior distributions. At $95\%$ credible level, the upper limits of $h^2\Omega_\text{pt}(25\;\text{Hz})$ are $1.5\times10^{-8}$, $1.3\times10^{-8}$, $1.4\times10^{-8}$ and $1.2\times10^{-8}$ for bubble collision dominant case fitted by envelope approximation, semi-analytic, numerical methods and sound wave dominant case, respectively.

\begin{figure}[ht]
    \centering
    \includegraphics[width=\columnwidth]{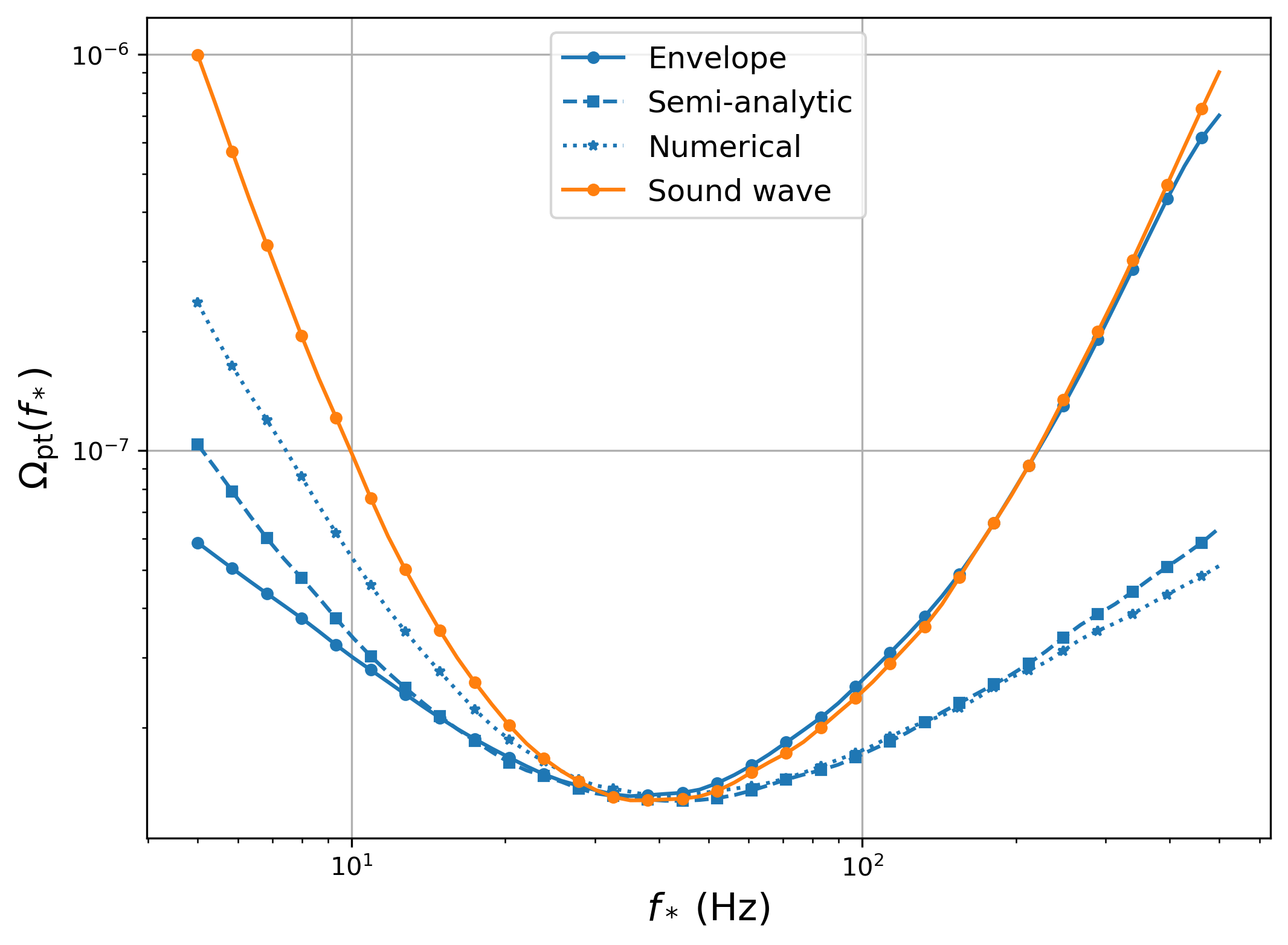}
    \caption{Upper limit of $\Omega(f_*)$ at $95\%$ credible level for certain values of peak frequency $f_*$.}
    \label{fig:upperlimit}
\end{figure}

In order to effectively demonstrate the constraints on the GW energy spectrum in the frequency band, we provide the upper limits on the amplitude of the GW energy spectrum $\Omega_\text{pt}(f_*)$ for the peak frequency $f_*$ in the range of $f_*\in [5,500]$ Hz. Our results are illustrated in \Fig{fig:upperlimit}, 
 and we find that the constraints are different for these four different theoretical models. However, the most stringent constraint appears at around $f_*\simeq40\;\text{Hz}$ with the upper limits of $\Omega_\text{pt}(f_*\simeq40\;\text{Hz})<1.3\times10^{-8}$ at $95\%$ credible level for all four scenarios. 
\smallskip

{\it Discussion and Conclusions.}
We search for the GW signals generated by first-order PT in various theoretical models widely adopted in literature using data from Advanced LIGO and Advanced Virgo's first three observing runs. To demonstrate the influence of theoretical uncertainties, three models of bubble collision and one of sound wave contribution are investigated separately. The result of Bayesian analysis does not show preference between different models and implies that it is impossible to tell which model is more preferred. In all, we find that, roughly speaking, $H_*/\beta\lesssim 0.1$ and $\alpha\lesssim 1$ at $68\%$ credible level in the peak frequency range of $20\lesssim f_*\lesssim 100$ Hz corresponding to the most sensitive frequency band of Advanced LIGO and Advanced Virgo's first three observing runs.

Generally, a log-uniform prior is preferred for variables spanning several orders of magnitude. However, the lower bound of a log-uniform distribution cannot be taken to be zero, and how to choose such a specific value is unclear. Because there is no correlation found in the raw data, the result of posterior depends on the choice of priors. The initial sampling points are mostly concentrated in small values of parameters for log-uniform priors and might lead to an under-estimation of the upper bound of the GW energy spectrum. In this sense, we adopt uniform prior distributions to avoid ambiguities in the choice of priors.
Since there is no correlated signal examined from the observing periods, we put the upper limits on the amplitude of the GW energy spectrum in the peak frequency range of $f_*\in [5,500]$ Hz. Because the GW spectra of the four theoretical models considered in this letter are roughly the same around the peaks, the constraints on the amplitudes of the GW spectra are roughly the same for them if the peak frequency of the GW energy spectra stays within the most sensitive frequency band of Advanced LIGO and Advanced Virgo, such as $f_*\in [20,80]$ Hz. But, once the peak frequency stays outside the the most sensitive frequency band, the shape of GW spectra plays an important role on the data analysis and the constraints should be different for different theoretical models. The results in \Fig{fig:upperlimit} are consistent with what we expect.

%Although conservative upper limits  have been led by using uniform priors, we have to point out that \Eq{eq:template} may be invalid to fit a strong phase transition, which limits the range of parameters. We find that the upper limits of $h^2\Omega$ at $25\,\text{Hz}$ varies from $1.2\times10^{-8}$ to $1.5\times10^{-8}$ for different models, which is an insignificant difference. Then we report the constrain curves on the frequency band in \Fig{fig:upperlimit}. Again, the limitations of different models in the frequency domain have no obvious differences in the sensitivity range of the detectors and the strongest limit occurs at about $40\,\text{Hz}$. (*???*)

\smallskip
{\it Acknowledgments. } 
We would like to thank Li Li, Jiang-Hao Yu and Yue Zhao for their useful conversations. 
We acknowledge the use of HPC Cluster of ITP-CAS and HPC Cluster of Tianhe II in National Supercomputing Center in Guangzhou. This work is supported by the National Key Research and Development Program of China Grant No.2020YFC2201502, grants from NSFC (grant No. 11975019, 11991052, 12047503), Key Research Program of Frontier Sciences, CAS, Grant NO. ZDBS-LY-7009, CAS Project for Young Scientists in Basic Research YSBR-006, the Key Research Program of the Chinese Academy of Sciences (Grant NO. XDPB15).

%%%%%%%%%%%%%%%%%%%%%%%%%%%%%%%%%%%%%%%%%%%%%%%%%%%%%%%%%%%%%%%%%%%%%%%%%%%%%%%%
%%%%%%%%%%%%%%%%%%%%%%%%%%%%%%%%%%%%% references %%%%%%%%%%%%%%%%%%%%%%%%%%%%%%%%%%%%
%%%%%%%%%%%%%%%%%%%%%%%%%%%%%%%%%%%%%%%%%%%%%%%%%%%%%%%%%%%%%%%%%%%%%%%%%%%%%%%%
%\bibliographystyle{apj}
\bibliography{ref.bib}
	
	%%%%%%%%%%%%%%%%%%%%%%%%%%%%%%%%%%%%%%%%%%%%%%%%%%%%%%%%%%%%%%%%%%%%%%%%%%%%%%%%
\end{document}